\documentclass[prd,aps,
tightenlines,superscriptaddress,showpacs,
nofootinbib,eqsecnum,amsfonts,amsmath]{revtex4}
\usepackage{bm,graphics,graphicx,epsfig,color,ulem}
\allowdisplaybreaks

\def\gap{\gtrsim}
\def\be{\begin{equation}}
\def\ee{\end{equation}}
\def\ba{\begin{eqnarray}}
\def\ea{\end{eqnarray}}
\def\ud{\underline}
\def\la{\langle}
\def\ra{\rangle}
\def\bse{\begin{subequations}}
\def\ese{\end{subequations}}

\begin{document}
\title{2.5PN linear momentum flux from inspiralling compact binaries in 
quasicircular orbits and associated recoil: Nonspinning case}
\author{Chandra Kant Mishra}\email{chandra@rri.res.in}
\affiliation{Raman Research Institute, Bangalore 560 080, India}
\affiliation{Indian Institute of Science, Bangalore 560 012, India}
\author{K. G. Arun} \email{kgarun@cmi.ac.in} 
\affiliation{Chennai Mathematical Institute, Siruseri, 603103, India}
\author{Bala R. Iyer} \email{bri@rri.res.in}
\affiliation{Raman Research Institute, Bangalore 560 080, India}
\date{\today}
\begin{abstract}
Anisotropic emission of gravitational waves (GWs) from inspiralling 
compact binaries leads to the loss of linear momentum and hence 
gravitational recoil of the system. The loss rate of linear momentum 
in the far-zone of the source (a nonspinning binary system of black 
holes in quasicircular orbit) is investigated at the 2.5 post-Newtonian 
(PN) order and used to provide an analytical expression in harmonic 
coordinates for the 2.5PN accurate recoil velocity of the binary 
accumulated in the inspiral phase. We find that the recoil velocity at 
the end of the inspiral phase (i.e at the innermost stable circular 
orbit (ISCO)) is maximum for a binary with symmetric mass ratio 
of $\nu\sim0.2$ and is roughly about $\sim$4.58\,${\rm km\,s}^{-1}$. 
Going beyond inspiral, we also provide an estimate of the more important 
contribution to the recoil velocity from the plunge phase. Again the recoil 
velocity at the end of the plunge, involving contributions both from  inspiral 
and plunge phase, is maximum for a binary with $\nu\sim0.2$ and is of the 
order of $\sim$180\,${\rm km\,s}^{-1}$. 
\end{abstract}
\maketitle
\section{Introduction}
\label{sec:intro}
A coalescing black-hole (BH) binary which is anisotropically emitting gravitational waves (GWs) 
will experience a recoil as a consequence of the loss of linear momentum from the binary 
through outgoing GWs. This phenomenon of gravitational-wave recoil has substantial 
importance in astrophysics especially if one wants to study models which suggest the 
formation and growth of super massive black holes (SMBHs) at the centers of galaxies 
through successive mergers from smaller BHs (stellar or intermediate mass BHs)
\cite{Merritt:2004xa}. If the kick velocity of the product BH is more than its escape velocity 
from the host galaxy, the formation of SMBHs will not be favored, as would be the case with 
dwarf galaxies and globular clusters (see \cite{Komossa:2008qd} for observational evidence 
for ejection of the SMBHs). Even if the recoil velocity of the product BH is not sufficient 
to eject it from the host (which may be the case with giant elliptical galaxies), the product 
BH would be displaced from the center and eventually would fall back. Such a process may have 
important dynamical changes at the galactic core. 
For a more  detailed overview of  astrophysical possibilities,
 see Ref.\cite{Merritt:2004xa}. 
 One important claim of \cite{Merritt:2004xa} is that models 
that grow SMBHs from the mergers of smaller BHs will not be favored for the galaxies 
at redshifts $\rm{z}\,{\gap}\,10$ due to the difficulty in retaining the ``kicked'' black holes. 
However, observations of the local universe suggest 
that most of the galaxies (more than 50\% of them) have SMBHs at their centers 
\cite{Richstone98}. This shows that there must be something which prevents the ejection 
of the central black-hole from many of these galaxies and Ref.\cite{2007ApJ...667L.133S} 
investigates such questions. In the light of the above arguments 
it becomes important to have an accurate estimate of the recoil velocity of the 
coalescing binary black holes (BBHs).     

One of the first proposals investigating the phenomenon of gravitational-wave recoil
is due to Peres \cite{1962PhRv..128.2471P}. Following the analogy from the classical 
electrodynamics, it was suggested that the lowest order secular effects related to 
gravitational-wave recoil arise due to the interaction of mass quadrupole moment with 
mass octupole moment or current quadrupole moment. His work provided the first formal theory 
for gravitational-wave recoil of a general material system in linearized gravity and is 
valid for any kind of motion (rotational, vibrational or any other kind) given the source 
is localized within a finite volume. In another early work, Bonnor and Rotenberg 
\cite{BoR61} studied the emission of gravitational waves from a pair of oscillating particles 
and suggested the possibility of GW recoil. Papapetrou \cite{Papa71} derived the leading order 
formula which involved interaction of mass quadrupole moment with mass octupole 
moment and current quadrupole moment. Later, Thorne \cite{Th80} generalized the idea by 
providing a general multipole expansion for the linear momentum loss as seen in the far 
zone of the source.\\

Within the post-Newtonian (PN) scheme, the leading order contribution to the linear momentum flux 
from an inspiralling binary system of two point masses in Keplarian orbit was 
computed by Fitchett \cite{Fitchett83} and binary motion in circular orbit was 
discussed as a limiting case of the main results of the work. The first PN correction 
was added to it by Wiseman \cite{Wi92} and the circular orbit case was discussed as a 
special case. In a work by Blanchet, Qusailah and Will (hereafter BQW) \cite{BQW05} 
the 1PN expression for linear momentum flux, from a nonspinning compact binary moving in 
quasicircular orbit, was extended to the 2PN order by adding the
hereditary contribution that occurs at 1.5PN order and the instantaneous one occurring 
at the 2PN order. Kidder \cite{K95} computed the leading order (spin-orbit)
contribution to linear momentum flux for generic orbits and discussed circular orbit 
effects as a limiting case of his findings. Recently, Racine et al. \cite{Racine:2008kj} 
extended Kidder's work by adding higher order spin corrections (spin-orbit terms at 
1.5PN order, spin-orbit tail and spin-spin terms at 2PN order). They provided 2PN 
accurate expression for the linear momentum flux from spinning BBHs in generic orbits. 
They also specialize to the binary motion in circular orbit and provide
estimates for the recoil velocity, accumulated during the inspiral phase, for equal mass 
binaries with spins equal in magnitude but opposite in direction.   

Using the black-hole perturbation theory, Favata et al. \cite{Favata:2004wz} 
estimated the recoil velocity of the binary, treating it as 
a test particle inspiralling into a black-hole (spinning or nonspinning), up to the 
innermost stable circular orbit (ISCO) (accounting for the recoil velocity accumulated during 
the inspiral phase) very accurately. Though their calculations were valid only in extreme 
mass ratio limit ($q_{\rm m}\equiv m_1/m_2<<1$) they extrapolated their results to 
$q_{\rm m}\sim0.4$ (with modest accuracy) using some scaling results from quadrupole 
approximations. A crude estimate of the contributions due to the plunge was also given. 
Within the validity of the approach, their estimates suggested the typical recoil velocity 
can be of the order of 10-100\,${\rm km\,s}^{-1}$ but for some configurations it may reach
roughly up to 500\,${\rm km\,s}^{-1}$. Another computation by Damour and Gopakumar \cite{Damour:2006tr} used 
the effective one-body approach \cite{BuonD98,D01} to compute the total recoil velocity of 
the final black-hole taking into account the contributions from all the three phases 
(inspiral, plunge and ringdown). Depending upon the method they used to compute linear 
momentum flux their estimates for maximum recoil velocity lie in the range 
49-172\,${\rm km\,s}^{-1}$. 
Reference \cite{Sopuerta:2006et} presents estimates of the recoil velocities for binaries 
in orbits with small eccentricities using an approximation technique that is valid only 
for late stages of the plunge. They also combine their results with the PN estimates of 
recoil velocity at ISCO of BQW in order to give estimates for recoil velocity for binaries 
in quasicircular orbits and find that for a binary with symmetric mass ratio (ratio of reduced mass of 
the binary to the total mass) $\nu=0.2$ the recoil velocity 
estimates should lie in the range of (79-216) ${\rm km\,s}^{-1}$.
In a recent work \cite{LeTiec:2009yg}, the 
recoil of the final BH was investigated combining 
the results of \cite{BQW05} with the calculation of contribution from the ringdown phase 
performed using the close-limit approximation. They found that the radiation emitted in the 
ringdown phase produces a significant antikick and thus brings down the estimates of 
recoil velocity based on only inspiral and plunge phase, {\it e.g.} after including 
the contributions from ringdown phase the maximum recoil velocity of the final black-hole 
is of the order of 180\,${\rm km\,s^{-1}}$ as compared to BQW estimate of 
243\,${\rm km\,s^{-1}}$ which does not include the contribution from the ringdown phase 
(also see Fig.1 of \cite{LeTiec:2009yg} for a 
comparison of this result with various numerical and analytical estimates). In another recent work 
\cite{Sundararajan:2010sr}, the phenomenon of recoil of a spinning BBH (extreme mass ratio) 
due to the inspiral, merger and ringdown phase of its evolution has been investigated. 
The issue of antikick has been examined very carefully and they found that for orbits 
aligned with the BH spin, the antikick grows with the spin. Also, a prograde coalescence 
of a smaller BH into the rapidly rotating bigger BH results in the smallest kick, whereas the 
retrograde coalescence insures the maximum recoil.     

In addition to the analytical or semianalytical estimates of the recoil there have 
been many investigations using numerical techniques. Recent numerical simulations for 
nonspinning \cite{Campanelli:2004zw, Baker:2006vn, Herrmann07,
Gonzalez:2008bi} BBHs in quasicircular orbit have shown that the recoil velocity 
can reach up to a few hundred ${\rm km\,s}^{-1}$, while for the spinning case 
\cite{Herrmann:2007ac, Koppitz:2007ev, Campanelli:2007ew, Gonzalez:2007hi} the recoil velocity 
estimates are much higher and can be of the order of few thousand ${\rm km\,s}^{-1}$. 
Although numerical simulations can put better constraints on these estimates, 
such simulations (especially those which include BH spins) are computationally very 
expensive. Moreover a very detailed multipolar study of numerical results for BBH recoil 
\cite{Schnittman:2007ij} shows the need of analytical and semianalytical 
schemes in order to gain a deeper understanding of the problem at hand and also as a check to 
numerical results.     
 
In the present work we extend the 2PN calculation of \cite{BQW05} for linear momentum 
loss from a nonspinning BBH in quasicircular orbit by adding terms (both 
instantaneous and hereditary) which contribute at 2.5PN order and thus give an 
analytical expression for linear momentum flux which is now 2.5PN accurate.
Naturally, in the 2PN limit our expression for linear momentum flux given by 
Eq.~\eqref{eq:LMF} reduces to Eq.~(20) of \cite{BQW05}. The 2.5PN 
accurate expression for the recoil velocity of the binary is given by  
Eq.~\eqref{eq:recvel} which reduces to Eq.~(23) of \cite{BQW05} in the 2PN limit. 
For computing the contribution to the recoil velocity due to the plunge phase, we 
simply adopt the discussion given in Sec.~(4.1) of \cite{BQW05} and perform the 
computation using our 2.5PN accurate formulas. 
We find that the recoil velocity experienced by the binary (with $\nu=0.2$) 
at the end of inspiral (at fiducial ISCO) and end of the 
plunge (which includes the contributions from both inspiral and 
plunge phase) is roughly about $\sim$4.55\,${\rm km\,s}^{-1}$ and 
$\sim$179.5\,${\rm km\,s}^{-1}$, respectively.
In contrast, the recoil velocity at the end of the inspiral and the plunge 
using the 2PN formulas (see Fig.~1 of \cite{BQW05}) is of the order of 
22\,${\rm km\,s}^{-1}$ and 243\,${\rm km\,s}^{-1}$, respectively, corresponding 
to the same $\nu=0.2$. We see here that inclusion of 
terms at 2.5PN order brings down the estimates for the recoil velocity significantly,
exhibiting in this problem the feature arising from the asymptotic nature of the PN expansion 
and the need to explicitly investigate the next PN order implications of a calculation.
This also reminds us of a similar result of \cite{Wi92} where the inclusion of 1PN 
contribution brought down the Newtonian estimates since the 1PN term contributed negatively
to the recoil velocity. Something similar happens here and the large negative coefficients
at 2.5PN order (see Eq.~\eqref{eq:recvel}) brings down the 2PN estimates significantly.               
 
The paper is organized in the following way. In Sec.~\ref{sec:structure-LMF}, we first 
provide the PN structure of the linear momentum flux in terms of the radiative multipole 
moments and then we give explicit expressions for the instantaneous and hereditary 
contribution separately in terms of the source multipole moments. Section~\ref{sec:LMF}
starts with the formulas for source multipole moments with desired PN accuracy and 
next shows the computation of both instantaneous and hereditary contributions to the linear 
momentum in the far-zone of the binary. Finally, we give the 2.5PN accurate expression for 
the linear momentum flux by adding instantaneous and hereditary contributions. In 
Sec.~\ref{sec:recvel} we discuss the computation of the recoil velocity of the binary and 
also give the 2.5PN accurate analytical expression for the same. 
Section~\ref{sec:numerical-estimates} explores the method for  estimating the recoil 
velocity accumulated during the plunge phase. In Sec.~\ref{sec:results}, we present our 
numerical estimates of total recoil velocity and its dependence on the composition of the 
binary as well as final discussions.     
\section{The Post-Newtonian Structure for linear momentum flux}
\label{sec:structure-LMF}
The general formula for linear momentum flux in the far-zone of the source
in terms of symmetric trace-free (STF) radiative multipole moments
is given in \cite{Th80} and at relative 2.5PN order it takes the following form (see 
Eq.~(4.20\') of Ref.\cite{Th80}.) 
\begin{eqnarray}
\label{eq:structure-LMF}
{\mathcal{F}_{P}^{i}}(U)&=& \frac{G}{c^7}\,\biggl\{\left[\frac{2}{63}\,
U^{(1)}_{ijk}\,U^{(1)}_{jk}+\frac{16}{45}\,\varepsilon_{ijk}
U^{(1)}_{ja}\,V^{(1)}_{ka}\right]\nonumber\\&&
+{1\over c^2}\left[\frac{1}{1134}\,U^{(1)}_{ijkl}\,U^{(1)}_{jkl}
+\frac{1}{126}\,\varepsilon_{ijk}
U^{(1)}_{jab}\,V^{(1)}_{kab}
+\frac{4}{63}\,V^{(1)}_{ijk}\,V^{(1)}_{jk}\right]\nonumber\\&&
+{1\over
c^4}\left[\frac{1}{59400}\,U^{(1)}_{ijklm}\,U^{(1)}_{jklm}+\frac{2}{14175}\,\varepsilon_{ijk}
U^{(1)}_{jabc}\,V^{(1)}_{kabc}+
\frac{2}{945}\,V^{(1)}_{ijkl}\,V^{(1)}_{jkl}\right]+{\cal O}\left({1\over c^6}\right) \biggr\}.
\end{eqnarray}

In the above expression $U_{L}$ and $V_{L}$(where $L=i_1i_2\cdots i_l$
represents a multi-index composed of $l$ spatial indices) are the mass-type 
and current-type radiative multipole moments respectively and
$U_{L}^{(p)}$ and $V_{L}^{(p)}$ denote their $p^\mathrm{th}$ time
derivatives. The Levi-Civita tensor is denoted by $\varepsilon_{ijk}$, such 
that $\varepsilon_{123}=+1$ and ${\cal O}(1/c^6)$ indicates that we ignore 
contributions of the order 3PN and higher. The moments appearing in the formula 
are functions of retarded time $U\equiv T-({\it R}/c)$ in radiative coordinates.
Here $R$ and $T$ denote the distance of the source from the observer and the time 
of observation in radiative coordinates, respectively. 

It should be evident from Eq.~\eqref{eq:structure-LMF} that the computation of 
2.5PN accurate linear momentum flux requires the knowledge of $U_{ij}, V_{ij}$ 
and $U_{ijk}$ at 2.5PN order, $U_{ijkl}$ and $V_{ijk}$ at 1.5PN order, and 
$U_{ijklm}$ and $V_{ijkl}$ at Newtonian order. In a recent work \cite{BFIS08}, 
$U_L$ and $V_L$ have been computed with accuracies sufficient for the present 
purpose using multipolar post-Minkowskian (MPM) approximation approach 
\cite{BFeom, Bliving, BIJ02, BI04, BDE04, BDEI05}. In the 
multipolar post-Minkowskian formalism $U_{L}$ and $V_{L}$ are related to canonical moments $M_L$ and 
$S_L$ (Eqs. (5.4)-(5.8) of \cite{BFIS08}) which in turn are related to source 
moments $\left\{I_L, J_L, X_L, W_L, Y_L, Z_L\right\}$ (Eqs. (5.9)-(5.11) of \cite{BFIS08}). 
Rewriting the expressions for the radiative moments in terms of the source 
moments, the linear momentum flux can be decomposed as the 
sum of two distinct parts: the instantaneous terms and the hereditary terms.
By instantaneous we refer to contributions in the linear momentum flux
which depend on the dynamics of the system only at the retarded instant
$U\equiv T-({\it R}/c)$. Hereditary contributions to the flux, on the other hand,
are terms nonlocal in time depending on the dynamics of the system in its
entire past~\cite{BD92}. The linear momentum flux thus is conveniently 
decomposed into
\begin{eqnarray}
{\mathcal{F}_{P}^{i}}&=&\left({\mathcal{F}_{P}^{i}}\right)_{\rm inst}
+\left({\mathcal{F}_{P}^{i}}\right)_{\rm hered}\,,
\label{eq:decomposedLMF}
\end{eqnarray}
where the instantaneous part is given by 
\begin{eqnarray}
\label{eq:LMF-inst-IJ}
\left({\mathcal{F}_{P}^{i}}\right)_{\rm inst}&=&
\frac{G}{c^7}\,\left\{\frac{2}{63}\,
I^{(4)}_{ijk}\,I^{(3)}_{jk}+\frac{16}{45}\,\varepsilon_{ijk}
I^{(3)}_{ja}\,J^{(3)}_{ka}
\right.\nonumber\\&&\left.
+{1\over c^2}\left[\frac{1}{1134}\,I^{(5)}_{ijkl}\,I^{(4)}_{jkl}+
\frac{4}{63}\,J^{(4)}_{ijk}\,J^{(3)}_{jk}+\frac{1}{126}\,\varepsilon_{ijk}
I^{(4)}_{jab}\,J^{(4)}_{kab}\right]
\right.\nonumber\\&&\left.
+{1\over c^4}\left[\frac{1}{59400}\,I^{(6)}_{ijklm}\,I^{(5)}_{jklm}+
\frac{2}{945}\,J^{(5)}_{ijkl}\,J^{(4)}_{jkl}+\frac{2}{14175}\,\varepsilon_{ijk}
I^{(5)}_{jabc}\,J^{(5)}_{kabc}\right]
\right.\nonumber\\&&\left.
+{G\over c^5}\left[{2\over63}\left(I_{ijk}^{(4)}\left[{1\over7}\,I_{a\la j}^{(6)}I_{k\ra a}
-{4\over7}\,I_{a\la j}^{(5)}I_{k\ra a}^{(1)}
-I_{a\la j}^{(4)}I_{k\ra a}^{(2)}
-{4\over7}\,I_{a\la j}^{(3)}I_{k\ra a}^{(3)}
+{1\over3}\,\varepsilon_{ab\la j}I_{k\ra a}^{(5)}J_{b}
\right.\right.\right.\right.\nonumber\\&&\left.\left.\left.\left.
+4\left[W^{(2)}I_{jk}-W^{(1)}I_{jk}^{(1)}\right]^{(3)}\right]
+I_{jk}^{(3)}\left[
-{43\over12}\,I_{a\la i}^{(4)}I_{jk\ra a}^{(3)}
-{17\over12}\,I_{a\la i}^{(3)}I_{jk\ra a}^{(4)}
-3\,I_{a\la i}^{(5)}I_{jk\ra a}^{(2)}
\right.\right.\right.\right.\nonumber\\&&\left.\left.\left.\left.
+{1\over2}\,I_{a\la i}^{(2)}I_{jk\ra a}^{(5)}
-{2\over3}\,I_{a\la i}^{(6)}I_{jk\ra a}^{(1)}
+{1\over2}\,I_{a\la i}^{(1)}I_{jk\ra a}^{(6)}
+{1\over12}\,I_{a\la i}^{(7)}I_{jk\ra a}
+{1\over4}\,I_{a\la i}I_{jk\ra a}^{(7)}
+{1\over5}\,\varepsilon_{ab\la i}\left(
-12\,J_{j\ud{a}}^{(3)}I_{k\ra b}^{(3)}
\right.\right.\right.\right.\right.\nonumber\\&&\left.\left.\left.\left.\left.
-12\,I_{j\ud{a}}^{(3)}J_{k\ra b}^{(3)}
-15\,J_{j\ud{a}}^{(2)}I_{k\ra b}^{(4)}
-35\,I_{j\ud{a}}^{(2)}J_{k\ra b}^{(4)}
-4\,J_{j\ud{a}}^{(1)}I_{k\ra b}^{(5)}-36\,I_{j\ud{a}}^{(1)}J_{k\ra b}^{(5)}
-J_{j\ud{a}}I_{k\ra b}^{(6)}
-9\,I_{j\ud{a}}J_{k\ra b}^{(6)}
\right.\right.\right.\right.\right.\nonumber\\&&\left.\left.\left.\left.\left.
-{9\over4}\,J_{\ud{a}}I_{jk\ra b}^{(6)}
\right)
+{12\over5}\,J_{\la i}J_{jk\ra}^{(5)}
+4\,\left[W^{(2)}I_{ijk}-W^{(1)}I_{ijk}^{(1)}+3\,I_{\la ij}Y_{k\ra}^{(1)}\right]^{(4)}\right]\right)
+{16\over45}\,\varepsilon_{ijk}\left(I_{jp}^{(3)}
\right.\right.\right.\nonumber\\&&\left.\left.\left.
\left[{4\over7}\,J_{a\la k}^{(3)}I_{p\ra a}^{(3)}
+{8\over7}\,I_{a\la k}^{(3)}J_{p\ra a}^{(3)}
+3\,J_{a\la k}^{(2)}I_{p\ra a}^{(4)}+{5\over7}\,I_{a\la k}^{(2)}J_{p\ra a}^{(4)}
+{26\over7}\,J_{a\la k}^{(1)}I_{p\ra a}^{(5)}-{6\over7}\,I_{a\la k}^{(1)}J_{p\ra a}^{(5)}
+{9\over7}\,J_{a\la k}I_{p\ra a}^{(6)}
\right.\right.\right.\right.\nonumber\\&&\left.\left.\left.\left.
-{3\over7}\,I_{a\la k}J_{p\ra a}^{(6)}
-{1\over28}\,J_{a}I_{kpa}^{(6)}
-\varepsilon_{ab\la k}J_{\ud{a}}J_{p\ra b}^{(5)}
+{1\over14}\,\varepsilon_{ac\la k}
\left({425\over24}\,I_{p\ra bc}^{(3)}I_{ab}^{(4)}+{31\over12}\,I_{p\ra bc}^{(4)}I_{ab}^{(3)}
+{173\over6}\,I_{p\ra bc}^{(2)}I_{ab}^{(5)}
\right.\right.\right.\right.\right.\nonumber\\&&\left.\left.\left.\left.\left.
-{19\over24}\,I_{p\ra bc}^{(5)}I_{ab}^{(2)}
+{143\over8}\,I_{p\ra bc}^{(1)}I_{ab}^{(6)}
+{15\over4}\,I_{p\ra bc}I_{ab}^{(7)}+{3\over8}\,I_{p\ra bc}^{(7)}I_{ab}\right)
+2\left[\varepsilon_{ab\la k}\left(-I_{p\ra b}^{(3)}W_{a}-2\,I_{p\ra b}Y_{a}^{(2)}
\right.\right.\right.\right.\right.\right.\nonumber\\&&\left.\left.\left.\left.\left.\left.
+I_{p\ra b}^{(1)}Y_{a}^{(1)}\right)
+3\,J_{\la k}Y_{p\ra}^{(1)}-2\,J_{kp}^{(1)}W^{(1)}\right]^{(3)}\right]
+J_{kp}^{(3)}\left[{1\over7}\,I_{a\la j}^{(6)}I_{p\ra a}
-{4\over7}\,I_{a\la j}^{(5)}I_{p\ra a}^{(1)}
-I_{a\la j}^{(4)}I_{p\ra a}^{(2)}
-{4\over7}\,I_{a\la j}^{(3)}I_{p\ra a}^{(3)}
\right.\right.\right.\right.\nonumber\\&&\left.\left.\left.\left.
+{1\over3}\,\varepsilon_{ab\la j}I_{p\ra a}^{(5)}J_{b}
+4\left[W^{(2)}I_{jp}-W^{(1)}I_{jp}^{(1)}\right]^{(3)}\right]\right)
+{1\over1134}\,I_{jkl}^{(4)}\left(-20\,I_{\la ij}^{(3)}I_{kl\ra}^{(3)}
-{84\over5}\,I_{\la ij}^{(5)}I_{kl\ra}^{(1)}
-33\,I_{\la ij}^{(4)}I_{kl\ra}^{(2)}
\right.\right.\right.\nonumber\\&&\left.\left.\left.
-{21\over5}\,I_{\la ij}^{(6)}I_{kl\ra}
\right)
+{1\over126}\,\varepsilon_{ijk}\,I_{jpq}^{(4)}\left(
-{2\over5}\,\varepsilon_{ab\la k}I_{p\ud{a}}^{(5)}I_{q\ra b}^{(1)}
+{1\over10}\,\varepsilon_{ab\la k}I_{p\ud{a}}^{(6)}I_{q\ra b}
-{1\over2}\,\varepsilon_{ab\la k}I_{p\ud{a}}^{(4)}I_{q\ra b}^{(2)}
-2\,J_{\la k}I_{pq\ra}^{(5)}\right)
\right.\right.\nonumber\\&&\left.\left.
+{4\over63}\,J_{jk}^{(3)}\left(
-{2\over5}\,\varepsilon_{ab\la i}I_{j\ud{a}}^{(5)}I_{k\ra b}^{(1)}
+{1\over10}\,\varepsilon_{ab\la i}I_{j\ud{a}}^{(6)}I_{k\ra b}
-{1\over2}\,\varepsilon_{ab\la i}I_{j\ud{a}}^{(4)}I_{k\ra b}^{(2)}
-2\,J_{\la i}I_{jk\ra}^{(5)}
\right)
\right]
+{\cal O}\left({1\over c^6}\right)\right\},\nonumber\\ 
\end{eqnarray}
where,
\bse
\label{eq:M2M3S2}
\begin{align}
\left[W^{(2)}I_{ij}-W^{(1)}I_{ij}^{(1)}\right]^{(3)} &=
\left[2\,W^{(4)}I_{ij}^{(1)}+W^{(5)}I_{ij}-W^{(1)}I_{ij}^{(4)}-2\,W^{(2)}I_{ij}^{(3)}\right]
\label{eq:M2}\,,\\
\left[W^{(2)} I_{ijk}-W^{(1)} I_{ijk}^{(1)}+3\, I_{\la ij}Y_{k\ra}^{(1)}\right]^{(4)} &=
\left[W^{(6)}I_{ijk}
+3\,W^{(5)}\,I_{ijk}^{(1)}
+2\,W^{(4)}\,I_{ijk}^{(2)}
-3\,W^{(2)}\,I_{ijk}^{(4)}
\right.\nonumber \\&\left.
-2\,W^{(3)}\,I_{ijk}^{(3)}
-W^{(1)}I_{ijk}^{(5)}
+3\,I_{\la ij}Y_{k \ra}^{(5)}
+12\,I_{\la ij}^{(1)}Y_{k \ra}^{(4)}
\right.\nonumber \\&\left.
+18\,I_{\la ij}^{(2)}Y_{k \ra}^{(3)}
+12\,I_{\la ij}^{(3)}Y_{k \ra}^{(2)}
+3\,I_{\la ij}^{(4)}Y_{k\ra}^{(1)}
\right]
\label{eq:M3}\,,\\
\left[\varepsilon_{ab\la i}\left(-I_{j\ra b}^{(3)}W_{a}
-2\,I_{j\ra b}Y_{a}^{(2)}+I_{j\ra b}^{(1)}Y_{a}^{(1)}\right)
\right]^{(3)} &=\varepsilon_{ab\la i}
\left(
-I_{j\ra b}^{(6)}W_{a}
-3\,I_{j\ra b}^{(5)}W_{a}^{(1)}
-3\,I_{j\ra b}^{(4)}W_{a}^{(2)}
-I_{j\ra b}^{(3)}W_{a}^{(3)}
\right.\nonumber \\&\left.
-2\,I_{j\ra b}Y_{a}^{(5)}
-5\,I_{j\ra b}^{(1)}Y_{a}^{(4)}
-3\,I_{j\ra b}^{(2)}Y_{a}^{(3)}
+I_{j\ra b}^{(3)}Y_{a}^{(2)}
+I_{j\ra b}^{(4)}Y_{a}^{(1)}
\right)
\label{eq:S2a}\,,\\
\left[3\,J_{\la i}Y_{j\ra}^{(1)}-2\,J_{ij}^{(1)}W^{(1)}\right]^{(3)}&=
\left[
3\,J_{\la i}Y_{j\ra}^{(4)}
-2\,J_{ij}^{(1)}W^{(4)}
-6\,J_{ij}^{(2)}W^{(3)}
-6\,J_{ij}^{(3)}W^{(2)}
\right.\nonumber \\&\left.
-2\,J_{ij}^{(4)}W^{(1)}
\right]\,.
\label{eq:S2b}
\end{align}
\ese
In the above, angular brackets ($\la\ra$) surrounding indices 
denote symmetric trace-free projections. Underlined indices 
denote indices that are excluded in the symmetric trace-free 
projection. The hereditary contribution can 
be written as
\begin{eqnarray}
\label{eq:LMF-hered-IJ}
{\left({\mathcal{F}_{P}^{i}}\right)_{\rm hered}}&=&
\frac{4\,G^2\,M}{63\,c^{10}}\,I_{ijk}^{(4)}(U)
\int_{0}^{\infty} d\tau \left[\ln \left({\tau\over
2\tau_0}\right)+{11\over12}\right]
I^{(5)}_{jk}(U-\tau)
\nonumber\\&&
+\frac{4\,G^2\,M}{63\,c^{10}}\,I_{jk}^{(3)}(U)\int_{0}^{\infty} d\tau \left[\ln
\left({\tau\over2\tau_0}\right)+{97\over60}\right]
I^{(6)}_{ijk}(U-\tau)
\nonumber\\&&
+{32\,G^2\,M\over 45\,c^{10}}\,\varepsilon_{ijk}\,
I_{ja}^{(3)}(U)\int_{0}^{\infty} d\tau \left[\ln \left({\tau\over
2\tau_0}\right)+{7\over6} \right]
J^{(5)}_{ka}(U-\tau)
\nonumber\\&&
+{32\,G^2\,M\over 45\,c^{10}}\,\varepsilon_{ijk}\,J_{ka}^{(3)}(U)\int_{0}^{\infty} d \tau
\left[\ln \left({\tau\over 2\tau_0}\right)+{11\over12} \right]
I^{(5)}_{ja}(U-\tau)
\nonumber\\&&
+{G^2\,M\over 567\,c^{12}}\,
I_{ijkl}^{(5)}(U)\int_{0}^{\infty} d\tau \left[\ln \left({\tau\over
2\tau_0}\right)+{97\over60} \right]
I^{(6)}_{jkl}(U-\tau)
\nonumber\\&&
+{G^2\,M\over 567\,c^{12}}\,I_{jkl}^{(4)}(U)\int_{0}^{\infty} d \tau
\left[\ln \left({\tau\over 2\tau_0}\right)+{59\over30} \right]
I^{(7)}_{ijkl}(U-\tau)
\nonumber\\&&
+{G^2\,M\over 63\,c^{12}}\,\varepsilon_{ijk}\,I_{jab}^{(4)}(U)\int_{0}^{\infty} d\tau 
\left[\ln \left({\tau\over
2\tau_0}\right)+{5\over3} \right]
J^{(6)}_{kab}(U-\tau)
\nonumber\\&&
+{G^2\,M\over 63\,c^{12}}\,\varepsilon_{ijk}\,J_{kab}^{(4)}(U)\int_{0}^{\infty} d \tau
\left[\ln \left({\tau\over 2\tau_0}\right)+{97\over60} \right]
I^{(6)}_{jab}(U-\tau)
\nonumber\\&&
+{8\,G^2\,M\over 63\,c^{12}}\,J_{ijk}^{(4)}(U)\int_{0}^{\infty} d\tau 
\left[\ln \left({\tau\over
2\tau_0}\right)+{7\over6} \right]
J^{(5)}_{jk}(U-\tau)
\nonumber\\&&
+{8\,G^2\,M\over 63\,c^{12}}\,J_{jk}^{(3)}(U)\int_{0}^{\infty} d \tau
\left[\ln \left({\tau\over 2\tau_0}\right)+{5\over3} \right]
J^{(6)}_{ijk}(U-\tau)
+{\cal O}\left({1\over c^6}\right).
\end{eqnarray}
Here, $M$ denotes the ADM mass of the system. $\tau_0$ 
appearing in above hereditary integrals is an arbitrary constant and 
is related to an arbitrary length scale, $r_0$, by the relation 
$\tau_0$=$r_0/c$. It enters the relation connecting retarded time 
$U\equiv T-(R/c)$ in radiative coordinates to retarded time 
$u\equiv t_{\rm h}-r_{\rm h}/c$ in harmonic coordinates (where 
$r_{\rm h}$ is the distance of the source in harmonic coordinates). 
The relation between retarded time in radiative coordinates, 
and the one in harmonic coordinates reads as
\be
\label{eq:U-u}
U=t_{\rm h}-{r_{\rm h}\over c}-{2\,G\,M\over c^3}\,\log{\left({r_{\rm h}\over r_0}\right)}\,.
\ee
\section{The 2.5PN linear momentum flux: Application to Inspiralling 
compact binaries in circular orbits}
\label{sec:LMF}
Equations~\eqref{eq:decomposedLMF}-\eqref{eq:LMF-hered-IJ} collectively 
give the far-zone linear momentum flux from generic PN sources in terms
of the source multipole moments $\left\{I_L, J_L, W_L, Y_L\right\}$. In 
this section, we specialize to the case of nonspinning inspiralling compact 
binaries, in which two compact objects (neutron stars and/or black holes) are 
moving around each other in quasicircular orbits. All the source multipole 
moments in case of nonspinning inspiralling compact binaries moving in 
quasicircular orbits are now known with the accuracies sufficient for the present 
purpose and have been computed and listed in \cite{BFIS08} (see Eqs. (5.12)-(5.25) there). 
Here, we just quote those results with the accuracies that is required for the present 
purpose. For mass-type moments, we have
\bse
\label{eq:I2I3I4I5}
\begin{align}
I_{ij} &= \nu\,m\,\left\{{x}_{\la ij\ra}\left[1 + \gamma
\left(-{1\over 42}-{13\over 14}\nu \right) + \gamma^2 \left(-{461\over
1512} -{18395\over 1512}\nu - {241\over 1512} \nu^2\right)\right]
\right.\nonumber\\ &~~\left. +{r^2\over c^2}{v}_{\la ij\ra}\left[{11\over
21}-{11\over 7}\nu + \gamma \left({1607\over 378}-{1681\over 378} \nu
+{229\over 378}\nu^2\right)\right] + {48\over 7}\,{r\over c}\,\gamma^2\,\nu
\,x_{\la i}v_{j\ra}\right\}
+{\cal O}\left({1\over c^6}\right)\label{eq:I2}\,,\\
I_{ijk} &= -\nu\,m\,{\sqrt{1-4\,\nu}}\,\left\{ {x}_{\langle ijk\rangle} 
\left[1 -\gamma \nu+\gamma^2 \left(-{139\over 330}-{11923\over 660}\nu-{29\over
110}\nu^2\right) \right]\right.\nonumber\\ &~~+{r^2\over c^2}\,x_{\langle
i}v_{jk\rangle} \left[1 - 2\nu + \gamma \left({1066\over
165}-\left.{1433\over 330}\nu +{21\over 55} \nu^2\right)
\right]\right.\nonumber\\ &~~\left.+{196\over 15} \,{r\over
c}\,\gamma^2\,\nu\,x_{\langle ij}v_{k\rangle} \right\}
+{\cal O}\left({1\over c^6}\right)\label{eq:I3}\,,\\ 
I_{ijkl} &= \nu \,m\,\left\{ x_{\langle ijkl\rangle}\left[1 - 3\nu + \gamma
\left({3\over 110} - {25\over 22}\nu + {69\over22}\nu^2\right)\right]
\right.\nonumber\\&~~\left.
+{78\over55}{r^2\over c^2}\,x_{\langle ij}v_{kl\rangle}(1-5\nu+5\nu^2)\right\}
+{\cal O}\left({1\over c^4}\right)\label{eq:I4}\,,\\
I_{ijklm}&=-\nu\,m\,{\sqrt{1-4\,\nu}}\,x_{\la ijklm\ra}\left(1-2\nu\right)
+{\cal O}\left({1\over c^2}\right)\label{eq:I5}\,,
\end{align}
\ese
and, for the current-type moments
\footnote{The coefficient ``-{484/105}'' appearing at the 2.5PN order in the 
expression for $J_{ij}$ in Eq.~5.15b of \cite{BFIS08} is incorrect and should 
be replaced by ``-{188/35}'' (see \eqref{eq:J2} above and the erratum of \cite{BFIS08}).} we have
\bse
\label{eq:J2J3J4}
\begin{align}
J_{ij} &= -\nu\,m\,{\sqrt{1-4\,\nu}}\,\left\{ \varepsilon_{ab\langle i} x_{j\rangle a}v_b 
\left[1 +\gamma\,\left({67\over 28}-{2\over 7}\nu \right)+\gamma^2\left({13\over 9} 
-{4651\over252}\nu -{1\over 168}\nu^2 \right)\right]\right.\nonumber\\
&~~\left.-{188\over 35} \,{r\over c}\,\gamma^2\,\nu\,\varepsilon_{ab\langle
i} v_{j\rangle a}x_b \right\} +{\cal O}\left({1\over c^6}\right)\,,\label{eq:J2}\\
J_{ijk} &= \nu\,m
\,\left\{ \varepsilon_{ab\langle i} x_{jk\rangle a} v_b \left[1 - 3\nu +
\gamma \left({181\over 90} - {109\over 18}\nu + {13\over
18}\nu^2\right)\right]\right.\nonumber\\
&~~+\left.{7\over45}{r^2\over c^2}\,\varepsilon_{ab\langle i}x_a v_{jk\rangle
b}\left(1-5\nu+5\nu^2\right)\right\}+{\cal O}\left({1\over c^4}\right)\,,\label{eq:J3}\\
J_{ijkl} &= -\nu\,m\,{\sqrt{1-4\,\nu}}\,\varepsilon_{ab\la i}x_{jkl\ra a}v_b
\left(1-2\nu\right)+{\cal O}\left({1\over c^2}\right)\,.\label{eq:J4}
\end{align}
\ese 
Computation of linear momentum flux at 2.5PN order also requires, 1PN accurate 
expression for mass monopole moment, which can be identified with ADM mass ($M$) 
of the source, and Newtonian accurate expression for the current dipole moment 
$J_i$. We have 
\bse
\label{eq:IJ1}
\begin{align}
I&=M =m\left(1-{\nu\over2}\,\gamma\right)+{\cal O}\left({1\over c^4}\right)\,,\label{eq:I}\\
J_i &=\nu\,m\,\varepsilon_{abi}\,x_{a}\,v_{b}+{\cal O}\left({1\over c^2}\right)\,.\label{eq:J1}
\end{align}
\ese 
In addition to mass-type and current-type moments we also need some of the gauge 
moments which only need to be Newtonian accurate and are given as
\bse
\label{eq:WY}
\begin{align}
W &= {\cal O}\left({1\over c^2}\right)\label{eq:W0},\\ 
W_i &= {1\over 10} \,
\nu\,m \,{\sqrt {1-4\,\nu}}\, r^2\,v_i +
{\cal O}\left({1\over c^2}\right)\label{eq:W1},\\
Y_i &= {1\over 5} \,
\frac{G\,m^2\,\nu}{r} \,{\sqrt {1-4\,\nu}}\,x_i +
{\cal O}\left({1\over c^2}\right).\label{eq:Y1}
\end{align}
\ese
In the above, $m=m_1+m_2$ is the total mass of the binary with $m_1$ and $m_2$ 
as the binary's component masses and $\nu$ is the symmetric mass ratio and is 
defined by the combination $({m_1\,m_2}/m^2)$. $x^i$ and $v^i$ denote the binary's 
relative separation and relative velocity of the two objects constituting 
the binary, respectively, and can be defined as $x^i=y_1^i-y_2^i$ and 
$v^i={\rm d}x^i/{\rm d}t=v_1^i-v_2^i$ (where ($y_1^i, y_2^i$) and ($v_1^i, v_2^i$) 
are positions and velocities of components of the binary).  
$\gamma$ is a PN parameter and is defined by the quantity $(G\,m/c^2\,r)$. 
\subsection{Instantaneous Terms}
\label{subsec:LMF-inst}
Equation~\eqref{eq:LMF-inst-IJ} is the general formula for the instantaneous part of 
the linear momentum flux from generic PN sources in terms of the source multipole moments 
$\left\{I_L, J_L, W_L, Y_L\right\}$. Computation of linear momentum flux involves computing 
time derivatives of the source multipole moments which in turn requires the knowledge of 
equations of motion with appropriate PN accuracy. Linear momentum flux computation at 2.5PN 
order will thus require 2.5PN accurate equations of motion\cite{BFIS08,K08}.

Let the $x$-$y$ plane be the orbital plane of the binary.\footnote{Since we are considering only 
nonspinning binary systems in quasicircular orbits, the motion will be in a fixed plane.} If 
$\phi$ denotes the orbital 
phase of the binary giving the direction of the unit vector, ${\bf \hat{n}}={\bf x}/r$, along the 
binary's relative separation, then   
\be
{\bf \hat{n}}={\cos{\phi}}\,{\bf \hat{e}_x}+{\sin{\phi}}\,{\bf \hat{e}_y}\,.
\label{eq:n-phi}
\ee
The binary's relative separation, velocity and acceleration are given by
\bse
\label{eq:xva}
\begin{align}
\mathbf{x} &= r \,\mathbf{\hat{n}},\label{eq:x}\\ 
\mathbf{v} &= \dot r \,\mathbf{\hat{n}}
+ r \,\omega \,\bm{\hat{\lambda}},\label{eq:v}\\ 
\mathbf{a} &= (\ddot{r} - r
\,\omega^2) \,\mathbf{\hat{n}} + (r \,\dot{\omega} + 2\,\dot{r} \,\omega)
\,\bm{\hat{\lambda}}.\label{eq:a}
\end{align}
\ese
where an over dot denotes a time derivative and $r=|{\bf x}|$ is the distance 
between the two objects in the binary. The orbital frequency $\omega$ is given 
by $\omega=\dot{\phi}$. The motion of the binary can be described by the rotating 
orthonormal triad $({\bf \hat{n}}, {\bf \hat{\lambda}}, {\bf \hat{e}_z})$ with 
${\bm \hat{\lambda}}={\bf \hat{e}_z}\times{\bf \hat{n}}$. 

Up to 2PN order, one can model the binary's orbit as exact circular orbit 
with $\dot{r}=\dot{\omega}=0$, but at 2.5PN order orbit of the binary decays 
due to radiation reaction forces and one must include the inspiral effects.
The leading order effect is computed using energy balance equation assuming 
that system is losing its orbital energy only through gravitational radiation.
At the 2.5PN order, for $\dot{r}$ and $\dot{\phi}$ we have
\bse
\label{eq:romdot}\begin{align}
\dot{r} &= - \frac{64}{5} \sqrt{\frac{G m}{r}}~\nu\,\gamma^{5/2} +
{\cal O}\left({1\over c^7}\right)\label{eq:rdot},\\ 
\dot{\omega} &=\frac{96}{5} \,\frac{G m}{r^3}\,\nu\,\gamma^{5/2}+
{\cal O}\left({1\over c^7}\right)\label{eq:omdot}.
\end{align}
\ese
By substituting the expressions for $\dot{r}$ and $\dot{\omega}$ 
in Eq.~\eqref{eq:v}-\eqref{eq:a} one can write for the relative inspiral
velocity and relative acceleration as
\bse
\label{va-insp}
\begin{align}
\mathbf{v} &= r \,\omega \,\bm{\hat{\lambda}} - \frac{64}{5}\sqrt{\frac{G
m}{r}}~\nu\,\gamma^{5/2}\,\mathbf{\hat{n}} +
{\cal O}\left({1\over c^6}\right)\label{eq:v-insp},\\ 
\mathbf{a} &= -\omega^2
\,\mathbf{x} - \frac{32}{5}\,\sqrt{\frac{G
m}{r^3}}\,\,\nu\,\gamma^{5/2}\,\mathbf{v} +
{\cal O}\left({1\over c^6}\right)\label{eq:a-insp}\,.
\end{align}
\ese
Finally, we give the PN expression for orbital frequency as a function of the 
binary's separation $r$ which is now known with 3PN accuracy \cite{BFIS08} 
but in the present work we just need the 2PN accurate expression. In harmonic 
coordinates it is given as 
\begin{align}
\label{eq:omega}
\omega^2 &= {G m\over r^3}\biggl\{ 1+\gamma\Bigl(-3+\nu\Bigr) + \gamma^2
\left(6+\frac{41}{4}\nu +\nu^2\right)+
{\cal O}\left({1\over c^6}\right)\biggr\}\,.
\end{align}
It is often convenient to use a parameter $x$ which is directly connected 
to the orbital frequency rather instead of using the PN parameter $\gamma$.
\footnote{The use of $x$ as a PN parameter is useful since 
it remains invariant for a large class of coordinate transformations including 
the harmonic and Arnowitt, Deser and Misner (ADM) coordinate systems.}
Our new parameter $x$ is related to 
orbital frequency (Eq. 6.5 in \cite{BFIS08}) as
\be
\label{eq:xomega}
x=\left({G\,m\,\omega\over c^3}\right)^{2/3}.
\ee  
A relation between $\gamma$ and $x$ can be obtained by using Eq.~\eqref{eq:xomega}
in Eq.~\eqref{eq:omega} and inverting for $\gamma$ in terms of $x$.
At the 2PN order, the PN parameter $\gamma$ is related to the parameter $x$ as
\begin{align}
\label{eq:gammax}
\gamma &= x \biggl\{ 1+x \left(1-\frac{\nu}{3}\right) + x^2
\left(1-\frac{65}{12}\nu\right) 
+{\cal O}\left({1\over c^6}\right)\biggr\}.
\end{align}
Now, we have all the inputs to compute the derivatives of source multipole moments 
with accuracies sufficient for the computation of 2.5PN accurate expression for 
linear momentum flux. Once we have the desired time derivatives of various source 
multipole moments, we can insert them in Eq.~\eqref{eq:LMF-inst-IJ} to get the 2.5PN 
accurate instantaneous part of the linear momentum flux. After a tedious but straightforward 
computation, we get for the 2.5PN accurate expression for linear momentum flux in terms of 
the parameter $\gamma$ 
\ba
\label{eq:LMF-inst-gamma}
{\left({\mathcal{F}_{P}^{i}}\right)_{\rm inst}}&=&
-{464\over105}\,{c^4\over G}\,\sqrt{1-4\,\nu}\,\gamma^{11/2}\,\nu^2
\left\{
\left[
1+\gamma\left(-{1861\over174}-{91\over261}\,\nu\right)
+\gamma^2\left({139355\over2871}+{36269\over1044}\,\nu+{17\over3828}\,\nu^2\right)
\right]\,{\hat{\lambda}_{i}}
\right.\nonumber\\&&\left.
+{3437\over870}\,\gamma^{5/2}\,\nu\,{\hat{n}_{i}}
+{\cal O}\left({1\over c^6}\right)\right\}.
\ea
Alternatively, we can rewrite the instantaneous part of linear momentum flux given by 
Eq.\eqref{eq:LMF-inst-gamma} in terms of the parameter $x$ by using 
Eq.~\eqref{eq:gammax} in the above equation. At 2.5PN order the linear momentum flux 
in terms of the parameter $x$ reads as 
\ba
{\left({\mathcal{F}_{P}^{i}}\right)_{\rm inst}}&=&
-{464\over105}\,{c^4\over G}\,\sqrt{1-4\,\nu}\,x^{11/2}\,\nu^2
\left\{
\left[
1+x\left(-{452\over87}-{1139\over522}\,\nu\right)
+x^2\left(-{71345\over22968}+{36761\over2088}\,\nu+{147101\over68904}\,\nu^2\right)
\right]\,{\hat{\lambda}_{i}}
\right.\nonumber\\&&\left.
+{3437\over870}\,x^{5/2}\,\nu\,{\hat{n}_{i}}
+{\cal O}\left({1\over c^6}\right)\right\}.
\label{eq:LMF-inst-x}
\ea 
\subsection{Hereditary Terms}
\label{subsec:LMF-hered}
In this subsection, we shall compute the hereditary contribution to linear momentum flux 
from a nonspinning inspiralling compact binary in quasicircular orbits, which in terms 
of the source multipole moments is given by Eq.~\eqref{eq:LMF-hered-IJ}. The leading order 
hereditary contribution (1.5PN term) for the nonspinning compact binaries in circular orbit have been 
computed in \cite{BQW05} and later confirmed by Racine et al.\cite{Racine:2008kj}. In this section 
we extend the computation of the hereditary contributions by adding terms contributing at the 
2.5PN order.

If the $x$-$y$ plane is the binary's orbital plane and the orbital phase at a given retarded time $U$ be 
$\phi({U})$ then unit vectors ${\bf \hat{n}}$ and $\hat{{\bm \lambda}}$ can be written as
\bse
\label{eq:nlphiU}
\begin{align}
{\bf \hat{n}}(U)&={\cos{\phi(U)}}\,{\bf \hat{e}_x}+{\sin{\phi(U)}}\,{\bf \hat{e}_y}\,,
\label{eq:n-phiU}\\
{\hat{\bm \lambda}}(U)&=-{\sin{\phi(U)}}\,{\bf \hat{e}_x}+{\cos{\phi(U)}}\,{\bf \hat{e}_y}\,.
\label{eq:l-phiU}
\end{align}
\ese
It is evident from Eq.~\eqref{eq:LMF-hered-IJ} that to compute the hereditary contribution 
one must know the relevant multipole moments and their derivatives both at any retarded time 
$U$ as well as at some other time $U'\equiv U-\tau<U$. Since multipole moments at retarded 
time $U'$ shall involve ${\bf \hat{n}}$ and $\hat{{\bm \lambda}}$ at $U'$, it would be useful 
to express ${\bf \hat{n}}(U')$ and $\hat{{\bm \lambda}}(U')$ in terms of ${\bf \hat{n}}(U)$ and 
$\hat{{\bm \lambda}}(U)$, which are independent of the integration variable, $\tau$, and thus one 
can pull out the vector quantities out side the hereditary integral. Following \cite{Racine:2008kj}, one 
possible way is to express ${\bf \hat{n}}(U')$ and $\hat{{\bm \lambda}}(U')$ as a linear 
combination of ${\bf \hat{n}}(U)$ and $\hat{{\bm \lambda}}(U)$ as
\bse
\label{eq:nlphiU1}
\begin{align}
{\bf \hat{n}}(U')&={\cos\left[{\phi(U)-\phi(U')}\right]}\,{\bf \hat{n}}-{\sin\left[{\phi(U)-\phi(U')}
\right]}\,\hat{{\bm \lambda}}\,,\label{eq:n-phiU1}\\
\hat{\bm \lambda}(U')&={\sin\left[{\phi(U)-\phi(U')}\right]}\,{\bf \hat{n}}
+{\cos\left[{\phi(U)-\phi(U')}\right]}\,{\hat{\bm \lambda}}\,.\label{eq:l-phiU1}
\end{align}
\ese
It should be evident from Eq.~\eqref{eq:LMF-hered-IJ} that hereditary contributions at 
the 2.5PN order require 1PN accuracy for the quantities appearing in first four terms 
while in remaining six terms they need  only be Newtonian accurate. It should be clear that 
while computing the time derivatives of the source multipole moments for hereditary 
contributions, the equations of motion need be only 1PN accurate at most. To start, let us consider 
the combinations of derivatives of source multipole moments appearing in the first term 
of Eq.~\eqref{eq:LMF-hered-IJ}, i.e $I_{ijk}^{(4)}(U)\,I_{jk}^{(5)}(U-\tau)$. We can 
use Eq.~\eqref{eq:nlphiU1} to express the quantity at hand in terms of ${\bf \hat{n}}(U)$ and 
$\hat{\bm \lambda}(U)$ and then perform the contraction of indices. After some straightforward algebra, we have
\begin{align}
\label{eq:T1}
I_{ijk}^{(4)}(U)\,I_{jk}^{(5)}(U')&={16\over5}\,\frac{c^{17}}{G^4\,m^2}\,x^{17/2}
\,\sqrt{1-4\,\nu}\,\nu ^2\Biggl\{\left[-203\,\sin(2\,{\delta{\phi}})+x\left(\frac{2657}{2}
\,\sin(2\,{\delta{\phi}})-\frac{1341}{2}\,\nu\,\sin(2\,{\delta{\phi}})\right)\right]
{\hat{n}_i(U)}\nonumber\\&+\left[202\,\cos(2\,{\delta{\phi}})+x\left(-\frac{9263}{7}\,\cos
(2\,{\delta{\phi}})+\frac{4689}{7}\,\nu\,\cos(2\,{\delta{\phi}})\right)\right]
{\hat{\lambda}_i(U)}\Biggr\}\,,
\end{align}
where we have defined ${\delta{\phi}}\equiv \phi(U)-\phi(U-\tau)$.\\
Similarly we can write for combinations of source multipole moments in 
various terms of Eq.~\eqref{eq:LMF-hered-IJ} as
\bse
\label{eq:T2toT10}
\begin{align}
I_{jk}^{(3)}(U)\,I_{ijk}^{(6)}(U')&={2\over5}\,\frac{c^{17}}{G^4\,m^2}\,x^{17/2}
\,\sqrt{1-4\,\nu}\,\nu ^2\Biggl\{\left[\sin({\delta{\phi}})+3645\,\sin(3\,{\delta
{\phi}})+x\left(-\frac{73}{14}\,\sin({\delta{\phi}})+\frac{9}{14}\,\nu\,\sin({\delta{\phi}})
\right.\right.\nonumber\\&\left.\left.-\frac{334125}{14}\,\sin(3\,{\delta{\phi}})
+\frac{168885}{14}\,\nu\,\sin(3\,{\delta{\phi}})\right)\right]{\hat{n}_i(U)}
+\left[-\cos({\delta{\phi}})+3645\,\cos(3\,{\delta{\phi}})+x\left(\frac{73}{14}
\,\cos(\,{\delta{\phi}})\right.\right.\nonumber\\&\left.\left.-\frac{9}{14}\,\nu
\,\cos({\delta{\phi}})-\frac{334125}{14}\,\cos(3\,{\delta{\phi}})+\frac{168885}{14}
\,\nu\,\cos(3\,{\delta{\phi}})\right)\right]{\hat{\lambda}_i(U)}
\Biggr\},\label{eq:T2}\\
{\varepsilon_{ijk}}\,I_{ja}^{(3)}(U)\,J_{ka}^{(5)}(U')&=
-2\,\frac{c^{17}}{G^4\,m^2}\,x^{17/2}\,\sqrt{1-4\,\nu}\,\nu ^2
\Biggl\{\left[-\sin({\delta{\phi}})+x\left(\frac{265}{84}\,\sin({\delta{\phi}})-\frac{85}{42}
\,\nu\,\sin({\delta{\phi}})\right)\right]{\hat{n}_i(U)}\nonumber\\&+\left[\cos({\delta{\phi}})
+x\left(-\frac{265}{84}\,\cos({\delta{\phi}})+\frac{85}{42}\,\nu\,\cos({\delta{\phi}})\right)
\right]{\hat{\lambda}_i(U)}\Biggr\},\label{eq:T3}\\
{\varepsilon_{ijk}}\,J_{ka}^{(3)}(U)\,I_{ja}^{(5)}(U')&=
{-8}\,\frac{c^{17}}{G^4\,m^2}\,x^{17/2}\,\sqrt{1-4\,\nu}\,\nu ^2\Biggl\{\left[\sin(2\,{\delta
{\phi}})+x\left(-\frac{265}{84}\,\sin(2\,{\delta{\phi}})+\frac{85}{42}\,\nu\,\sin(2\,{\delta
{\phi}})\right)\right]{\hat{n}_i(U)}\nonumber\\&+\left[\cos(2\,{\delta{\phi}})+x\,\left(
-\frac{265}{84}\,\cos(2\,{\delta{\phi}})+\frac{85}{42}\,\nu\,\cos(2\,{\delta{\phi}})\right)
\right]\,{\hat{\lambda}_i(U)}\Biggr\},\label{eq:T4}\\
I_{ijkl}^{(5)}(U)\,I_{jkl}^{(6)}(U')&=
{12\over7}\,\frac{c^{19}}{G^4\,m^2}\,x^{19/2}\,\sqrt{1-4\,\nu}\,(1-3\,\nu)\,\nu ^2\Biggl\{
\bigg[-\sin({\delta{\phi}})-54675\,\sin(3\,\,{\delta{\phi}})\bigg]{\hat{n}_i(U)}\nonumber\\&+
\bigg[\cos({\delta{\phi}})+54189\,\cos(3\,{\delta{\phi}})\bigg]\,{\hat{\lambda}_i(U)}
\Biggr\},\label{eq:T5}\\
I_{jkl}^{(4)}(U)\,I_{ijkl}^{(7)}(U')&=
{96\over7}\,\frac{c^{19}}{G^4\,m^2}\,x^{19/2}\,\sqrt{1-4\,\nu}\,(1-3\,\nu)\,\nu ^2\Biggl\{
\bigg[14\,\sin(2\,{\delta{\phi}})+12096\,\sin(4\,{\delta{\phi}})\bigg]\,{\hat{n}_i(U)}\nonumber\\
&+\bigg[-13\,\cos(2\,{\delta{\phi}})+12096\,\cos\,(4\,{\delta{\phi}})\bigg]\,{\hat{\lambda}_i(U)}
\Biggr\},\label{eq:T6}\\
{\varepsilon_{ijk}}\,I_{jab}^{(4)}(U)\,J_{kab}^{(6)}(U')&={32\over3}\,\frac{c^{19}}{G^4\,m^2}
\,x^{19/2}\,\sqrt{1-4 \nu}\,(1-3\,\nu)\,\nu^2\Biggl\{40\,\sin(2\,{\delta{\phi}})\,{\hat{n}_i(U)}
-41\,\cos(2\,{\delta{\phi}})\,{\hat{\lambda}_i(U)}\Biggr\},\label{eq:T7}\\
{\varepsilon_{ijk}}\,J_{kab}^{(4)}(U)\,I_{jab}^{(6)}(U')&=
{4\over3}\,\frac{c^{19}}{G^4\,m^2}\,x^{19/2}\,\sqrt{1-4 \nu}\,(1-3\,\nu)\,\nu^2\Biggl\{
\bigg[\sin({\delta{\phi}})-729\,\sin(3\,{\delta{\phi}})\bigg]{\hat{n}_i(U)}\nonumber\\&+\bigg[
-\cos({\delta{\phi}})-729\,\cos(3\,{\delta{\phi}})\bigg]{\hat{\lambda}_i(U)}\Biggr\},
\label{eq:T8}\\
J_{ijk}^{(4)}(U)\,J_{jk}^{(5)}(U')&=-{8\over3}\,\frac{c^{19}}{G^4\,m^2}\,x^{19/2}\,
\sqrt{1-4\,\nu}\,(1-3\,\nu)\,\nu^2\Biggl\{\sin({\delta{\phi}})\,{\hat{n}_i(U)}-\cos({\delta{\phi}})
\,{\hat{\lambda}_i(U)}\Biggr\},\label{eq:T9}\\
J_{jk}^{(3)}(U)\,J_{ijk}^{(6)}(U')&={32\over3}\,\frac{c^{19}}{G^4\,m^2}\,x^{19/2}
\,\sqrt{1-4\,\nu}\,(1-3\,\nu)\,\nu^2\Biggl\{\sin(2\,{\delta{\phi}})\,{\hat{n}_i(U)}
+\cos(2\,{\delta{\phi}})\,{\hat{\lambda}_i(U)}\Biggr\}.\label{eq:T10}
\end{align}
\ese

It is evident from the above that the dependence of the relevant quantities on the integration 
variable $\tau$ is only through $\delta{\phi}$ which under the assumption of adiabatic inspiral 
takes the form
\begin{align}
\label{dphi}
{\delta{\phi}}&={\phi(U)}-{\phi(U-\tau)}\nonumber\\
&={\phi(U)}-\bigg[\phi(U)-\tau\,\left(\frac{d\phi}{d\tau}\right)_{\tau=U}+\cdots
\bigg]\nonumber\\
&={\omega\,\tau}\,,
\end{align}
where second and higher derivatives of $\phi$ have been neglected.

Finally, one just needs the following standard integral to compute the hereditary terms in 
\eqref{eq:LMF-hered-IJ}
\be
\label{eq:stdint}
\int_0^\infty\log\left(\frac{\tau}{2\,b}\right)\,e^{i\,n\,{\omega}\,\tau}\,d\tau 
= {-\frac{1}{n\,{\omega}}}\left\{\frac{\pi}{2}{\rm Sign}\left[n\,\omega\right] + i\Big[\ln(2\,|n\,{\omega}|\,b) 
+{C} \Big]\right\}.
\ee

Equations \eqref{eq:T1}-\eqref{eq:stdint} provide all the necessary inputs that are 
needed for computing the hereditary terms. For the sake of compactness of the paper
we wish to skip some of the intermediate outcomes of the calculation and directly 
quote the final expression for the 2.5PN accurate hereditary contribution which in 
terms of the parameter $x$ reads as 
\ba
{\left({{\mathcal F}_{P}^i}\right)_{\rm hered}}&=&-{464\over105}\,{c^4\over G}
\,\sqrt{1-4\,\nu}\,x^{11/2}\,\nu^2\Biggl\{x^{3/2}\left[{309\over58}\,\pi\,{\hat{\lambda}_{i}}
+2\,\log\left({\omega\over {\omega}_{01}}\right){\hat{n}_{i}}\right]
+x^{5/2}\left[\left(-{2663\over116}\,\pi-{2185\over87}\,\pi\,\nu\right)\,{\hat{\lambda}_{i}}
\right.\nonumber\\&&\left.+\left(-\frac{106187}{50460}+\frac{32835}{841}\,{\log\,2}
-\frac{77625}{3364}\,{\log\,3}-{904\over87}\,\log\left(\omega\over{\omega}_{01}
\right)+\left[-\frac{38917}{25230}-\frac{109740}{841}\,{\log\,2}\right.\right.\right.
\nonumber\\&&\left.\left.\left.+\frac{66645}{841}\,{\log\,3}-{1400\over261}
\,\log\left(\omega\over{\omega}_{01}\right)\right]\,\nu\right){\hat{n}_{i}}\right]
+{\mathcal O}\left({1\over c^6}\right)\Biggr\},\label{eq:LMF-hered-x}
\ea 
where ${\omega_{01}}$ appearing in the above provides a scale to the logarithms 
and is given as
\ba
{\omega}_{01}&=&{1\over\tau_0}\,{\rm exp}\left({5921\over1740}+{48\over29}\,
{\log\,2}-{405\over116}{\log\,3}-C\right),
\label{omega0}
\ea
where $C$ is Euler's constant.
One can verify that terms involving the logarithms of frequency $\log\left({\omega\over
{\omega}_{01}}\right)$ appearing in Eq.~\eqref{eq:LMF-hered-x} can be reabsorbed into a 
new definition of phase variable and thus will disappear from the final expression for linear 
momentum flux. This possibility of introducing a new phase variable containing all the 
logarithms of frequency has been noticed and used in earlier works 
\cite{BIWW96, ABIQ04, BQW05}. We define the new phase variable $\psi$ as
\be
\label{modphase}
\psi=\phi-{2\,G\,M\,\omega\over c^3}\,{\log\left(\omega\over{\omega}_{01}\right)},
\ee
where $M$ is the ADM mass of the source and is given by Eq.~\eqref{eq:I}.
\subsection{Total LMF}
\label{subsec:TotalLMF}
The final expression for the LMF in terms of the parameter $x$ can be obtained by simply adding 
Eq.~\eqref{eq:LMF-inst-x} and Eq.~\eqref{eq:LMF-hered-x} and using $\psi$, given by Eq.~\eqref{modphase}, as the phase variable. In the final form the 2.5PN expression for LMF reads as
\ba
\label{eq:LMF}
{{\mathcal F}_{P}^i}&=&
-{464\over105}\,{c^4\over G}\,\sqrt{1-4\,\nu}\,x^{11/2}\,\nu^2
\left\{
\left[
1+x\left(-{452\over87}-{1139\over522}\,\nu\right)
+{309\over58}\,\pi\,x^{3/2}
+x^2\left(-{71345\over22968}+{36761\over2088}\,\nu
\right.\right.\right.\nonumber\\&&\left.\left.\left.
+{147101\over68904}\,\nu^2\right)
+x^{5/2}\left(-{2663\over116}\,\pi-{2185\over87}\,\pi\,\nu\right)
\right]\,{\hat{\lambda}_{i}}
+x^{5/2}\left[
-\frac{106187}{50460}+\frac{32835}{841}\,{\log\,2}
-\frac{77625}{3364}\,{\log\,3}
\right.\right.\nonumber\\&&\left.\left.
+\left(\frac{10126}{4205}
-\frac{109740}{841}\,{\log\,2}
+\frac{66645}{841}\,{\log\,3}\right)\,\nu
\right]{\hat{n}_i}+{\cal O}\left({1\over c^6}\right)
\right\}.
\ea
It should be clear that now ${\bf \hat{n}}$ and $\hat{\bm \lambda}$ are in the direction of new phase 
angle $\psi$ and $\psi+{\pi/2}$ respectively and are given as
\bse
\label{eq:nlpsi}
\begin{align}
{\bf \hat{n}}&={\cos{\psi}}\,{\bf \hat{e}_x}+{\sin{\psi}}\,{\bf \hat{e}_y}\,,
\label{eq:n-psi}\\
\hat{\bm \lambda}&=-{\sin{\psi}}\,{\bf \hat{e}_x}+{\cos{\psi}}\,{\bf \hat{e}_y}\,,
\label{eq:l-psi}
\end{align}
\ese
where $\psi$ is given by Eq.\eqref{modphase}.
\section{Recoil Velocity}
\label{sec:recvel}
Given the 2.5PN far-zone linear momentum flux due to a nonspinning inspiralling compact binary 
in quasicircular orbits (Eq.~\eqref{eq:LMF}) one can have 2.5PN accurate formula for the loss 
rate of linear momentum by the source using the linear momentum balance equation, which is 
\be
\label{eq:mombal}
{dP^i\over dt}=-{\mathcal F}_P^i\,.
\ee
The net loss of linear momentum can be obtained by integrating the balance equation, i.e.
\be
\label{eq:deltamom}
{\Delta P^i}=-\int_{-\infty}^{t}\,dt'\,{\mathcal F_P^i}\,.
\ee
For nonspinning compact objects moving in quasicircular orbit we have  
\bse
\label{eq:dndt-dldt}
\begin{align}
{d\hat{n}^i\over dt}&={\omega}\,{\hat{\lambda}^i}\,,\label{eq:dndt}\\
{d{\hat{\lambda}}^i\over dt}&=-{\omega}\,{\hat{n}^i}\,,\label{eq:dldt}
\end{align}
\ese
where $\omega$ is the orbital frequency of the inspiral. 
Computing the net change in the linear momentum (given by
Eq.~\eqref{eq:deltamom}) involves the following integrals
\begin{subequations}
\label{eq:n-l}
\begin{align}
\int_{-\infty}^{t}\omega^{11/3}\,{\hat{n}^i}\,dt'&=\int_{-\infty}^{t}\,{\omega^{8/3}}\,{d\hat{\lambda}^i\over dt'}\,dt'
=-{\omega^{8/3}}\,\Big[{\hat{\lambda}^i}-{8\over3}{\dot{\omega}\over {\omega^2}}\,\hat{n}_i\Big],\\
\int_{-\infty}^{t}\omega^{11/3}\,{\hat{\lambda}^i}\,dt'&=\int_{-\infty}^{t}\,{\omega^{8/3}}\,{d\hat{n}^i\over dt'}\,dt'
={\omega^{8/3}}\,\Big[{\hat{n}^i}+{8\over3}{\dot{\omega}\over {\omega^2}}\,\hat{\lambda}_i\Big].
\end{align}
\end{subequations}
Using Eq.~\eqref{eq:LMF} in Eq.~\eqref{eq:deltamom} and making use of integrals
given in Eq.~\eqref{eq:n-l} along with expressions for various relevant quantities listed in
Sec.~\ref{subsec:LMF-inst}, one can write the net change in linear momentum in terms of the PN parameter
$x$.\footnote{
Note that at the 2PN order, the net loss of linear
can be obtained by just replacing $\hat{n}^i$ by
$-\hat{\lambda}^i/\omega$ and $\hat{\lambda}^i$ by $\hat{n}^i/\omega$ in
Eq.~\eqref{eq:LMF} (see BQW \cite{BQW05} for a discussion). However, at the PN order of present
computations (2.5PN order) one needs to include the secular evolution of the orbital frequency $\omega$ and this has
been taken into account while writing Eq.~\eqref{eq:n-l}.} Once we have 
the net change in the momentum during the orbital evolution of the binary we can obtain the recoil velocity 
of the source by simply dividing it by the mass of the system i.e.      
\be
\label{v-dp}
{\Delta V}^i={{\Delta P}^i/m}\,,
\ee
and we find in terms of our parameter $x$, the 2.5PN accurate expression for the recoil velocity as
\begin{align}
\label{eq:recvel}
V_{\rm recoil}^{i}&={464\over105}\,c\,\sqrt{1-4\,\nu}\,x^4\,\nu^2\biggl\{\left[1+x\left(-{452\over87}-{1139\over522}\,\nu\right)
+{309\over58}\,\pi\,x^{3/2}+x^2\left(-{71345\over22968}+{36761\over2088}\,\nu+{147101\over68904}\,\nu^2\right)
\right.\nonumber\\&\left.
+x^{5/2}\left(-{2663\over116}\,\pi-{2185\over87}\,\pi\,\nu\right)\right]\,{\hat{n}_i}
+x^{5/2}\left[\frac{106187}{50460}-\frac{32835}{841}\,{\log\,2}+\frac{77625}{3364}\,{\log 3}+\left(\frac{41034}{841}+\frac{109740}{841}\,{\log\,2}
\right.\right.\nonumber\\&\left.\left.
-\frac{66645}{841}\,{\log\,3}
\right)\nu\right]{\hat{\lambda}_i}+{\cal O}\left({1\over c^6}\right)
\biggr\}.
\end{align}

\section{Numerical Estimates of Recoil velocity}
\label{sec:numerical-estimates}
With 2.5PN accurate formulas for the linear momentum flux (Eq.~\eqref{eq:LMF}) and
the recoil velocity (Eq.~\eqref{eq:recvel}), we now wish to compute the recoil 
velocity accumulated during the plunge phase. Generally, the PN approximation is considered to be less reliable 
for the orbits within the ISCO; by this we mean that PN corrections, when compared to 
the leading order contribution, become comparable. If these corrections are small 
even beyond the ISCO then one can use Eq.~\eqref{eq:LMF} to estimate  
the velocity accumulated during the plunge phase. Generally, it is 
expected that these corrections would become comparable to the leading order contribution 
near the common event horizon and thus one can only provide a crude estimate of the 
recoil velocity accumulated during the plunge.
For this purpose we simply adopt the methodology used in BQW \cite{BQW05} (see Sec.4.1 there\footnote{Though all 
necessary details have been given 
in \cite{BQW05} we provide here some of the basic formulas for completeness of the 
paper as well as for the convenience of the reader.}). We shall 
first compute the recoil velocity at the ISCO using Eq.~\eqref{eq:recvel}, where ISCO is 
taken to be that of a point particle moving around a Schwarzschild black-hole with the mass 
equal to the total mass of the binary i.e. $m=m_1+m_2$. For the kick velocity at the ISCO, we 
can write
\begin{align}
\label{eq:recvel-ISCO}
V_{\rm ISCO}^{i}&={464\over105}\,c\,\sqrt{1-4\,\nu}\,x_{\rm ISCO}^4\,\nu^2\biggl\{
\left[1+x_{\rm ISCO}\left(-{452\over87}-{1139\over522}\,\nu\right)+{309\over58}\,
\pi\,x_{\rm ISCO}^{3/2}+x_{\rm ISCO}^2\left(-{71345\over22968}+{36761\over2088}\,
\nu\right.\right.\nonumber\\&\left.\left.+{147101\over68904}\,\nu^2\right)
+x_{\rm ISCO}^{5/2}\left(-{2663\over116}\,\pi-{2185\over87}\,\pi\,\nu\right)\right]
\,{\hat{n}_{\rm ISCO}^i}+x_{\rm ISCO}^{5/2}\left[\frac{106187}{50460}-\frac{32835}{841}\,{\log\,2}
+\frac{77625}{3364}\,{\log 3}\right.\nonumber\\&\left.+\left(\frac{41034}{841}+\frac{109740}{841}\,{\log\,2}
-\frac{66645}{841}\,{\log\,3}
\right)\nu\right]\,{\hat{\lambda}_{\rm ISCO}^i}+{\mathcal O}\left({1\over c^6}\right)
\biggr\},
\end{align}
where $x_{\rm ISCO}$, $\hat{n}^i_{\rm ISCO}$ and $\hat{\lambda}^i_{\rm ISCO}$ denote 
values of $x, \hat{n}^i$ and $\hat{\lambda}^i$ at the ISCO respectively. For a 
point mass moving around a Schwarzschild black-hole of mass $m$ in circular orbits, 
$x_{\rm ISCO}$=1/6.
We choose for simplicity the phase at the ISCO to be $\psi=0$ and thus 
$\hat{n}_{\rm ISCO}^i=\{1,0,0\}$ and $\hat{\lambda}_{\rm ISCO}=\{0,1,0\}$. 
With this we can compute the recoil velocity due to the inspiral phase.\\

Following BQW, we adopt the effective one-body approach \cite{BuonD98,D01} to 
compute the plunge contribution to the recoil velocity. We assume that 
a point particle of mass $\mu$ is moving in the gravitational field 
of a Schwarzschild black-hole of mass $m$ where $\mu$ is the reduced mass of 
the system. In addition to this, we also ignore the effect of radiated energy 
and angular momentum on the plunge orbit. We have for the 
geodesic equations in the Schwarzschild geometry as
\bse
\label{eq:geodesiceqn}
\begin{align}
{dt\over d\tau}&={{\tilde{E}/c^2}\over 1-{2\,G\,m\over c^2\,r_{\rm s}}}\,,
\label{eq:dtdtau}\\
{d\psi\over d\tau}&={\tilde{L}\over r_{\rm s}^2}\,,\label{eq:dpsidtau}\\
{\left(dr_{\rm s}\over d\tau\right)^2}&={\tilde{E}^2/c^2}-c^2\left(1-{2\,G\,m\over c^2\,
r_{\rm s}}\right)\left(1+{\tilde{L}^2\over c^2\,r_{\rm s}^2}\right)\,.
\label{eq:drsdtau}
\end{align}     
\ese
Here, $\tau$ is the proper time along the geodesic and $\tilde{E}$ and 
$\tilde{L}$ are energy and orbital angular momentum per unit mass and 
can be defined in terms of dimensionless variables $\bar{E}$ and 
$\bar{L}$ as
\bse
\label{eq:EL}
\begin{align}
\tilde{E}&= c^2\,\bar{E}\,,\label{eq:tildeE}\\
\tilde{L}&={G\,m\over c}\,\bar{L}\,.\label{eq:tildeL}
\end{align}
\ese 
Using Eq.\eqref{eq:dpsidtau} and\eqref{eq:drsdtau}, one can obtain the phase 
of the orbit as
\be
\label{eq:psi-plng}
\psi=\int_{y_0}^{y}\,\left\{{\bar{L}\over \left[\bar{E}^2-(1-2\,y)
(1+\bar{L}^2\,y^2)\right]^{1/2}}\right\}\,dy,
\ee
with $y={(G\,m/r_{\rm s}\,c^2)}$ and for the phase at the beginning of the plunge 
we have chosen, $\psi=0$ (at $y=y_0$) to match the phase of the orbit at 
the ISCO.\\

Now the kick velocity accumulated during the plunge phase can be given by the 
following formula
\be
\label{eq:dVplng-t}
\Delta{V}_{\rm plunge}^i={1\over m}\,\int_{t_0}^{t_{\rm Horizon}}\,{dt\,{dP^i\over dt}}\,,
\ee
where $t_0$ and $t_{\rm Horizon}$ are the times at the beginning of the plunge and when 
the particle approaches the horizon, respectively.

Now it is clear from Eq.~\eqref{eq:dtdtau} that the time coordinate $t$ is singular in nature at the 
horizon, i.e. at $r_{\rm s}=2\,G\,m/c^2$, and thus we must have a variable which is nonsingular in nature 
at the horizon to compute the above integral. We can write Eq.~\eqref{eq:dVplng-t} as
\be
\label{eq:dVplng-om}
\Delta{V}_{\rm plunge}^i={1\over m}\,\int\,{\left(d{\bar{\omega}}\over d\bar{\omega}/dt\right){dP^i\over dt}}\,,
\ee
where the quantity, $\bar{\omega}$, is the proper angular frequency, defined as 
$\bar{\omega}={d\psi/d\tau}$, and is given by Eq.~\eqref{eq:dpsidtau}. After some straightforward algebra we
obtain
\be
\label{eq:dVplng-y}
\Delta{V}^i_{\rm plunge}={G\,\bar{L}\over c^3}\,\int_{y_0}^{y_{\rm Horizon}}\,\left({1\over x^{3/2}}
\,{dP^i\over dt}\right)\,\left({dy\over \left[\bar{E}^2-(1-2\,y)(1+\bar{L}^2\,y^2)
\right]^{1/2}}\right)\,,
\ee
where $dP/dt$ in terms of our parameter $x$ is given by Eq.~\eqref{eq:mombal} in combination with 
Eq.~\eqref{eq:LMF}. Now $x$ is related to the variable $y$ by the following equation and can be obtained 
after a few steps of algebra:
\be
\label{eq:xy}
x=\left[{\bar{L}\over \bar{E}}\,y^2\,(1-2\,y)\right]^{2/3}\,.
\ee
Using the above relation and the definition of phase given by Eq.~\eqref{eq:psi-plng}, 
the quantity inside the integral of Eq.~\eqref{eq:dVplng-y} becomes a function of just 
the integration variable, $y$. With known values of $\bar{E}$ and $\bar{L}$ for the 
plunge one can numerically compute the integral of Eq.~\eqref{eq:dVplng-y}. 

Our next task is to choose appropriate values for $\bar{E}$ and $\bar{L}$, which are also 
consistent with their values at the ISCO. In fact, there may be several ways to match a 
circular orbit at the ISCO to a suitable plunge orbit; we would use the two methods 
which have been used in \cite{BQW05}. In the first method, the particle is given an 
energy $\tilde{E}\equiv c^2\,\bar{E}$ such that, at the ISCO, and for an ISCO angular 
momentum $\tilde{L}_\mathrm{ISCO}=\sqrt{12}\,({G\,m/ c})$, its radial velocity is given 
by the standard quadrupole energy-loss formula for a circular orbit, which is given as
\be
\label{drhdt}
{dr_\mathrm{h}\over dt} =-{64\over 5}\left({G\,m\over c^2\,r_\mathrm{h}}\right)^3\nu\,c\,,
\ee
where $r_\mathrm{h}$ is the binary's radial separation in harmonic coordinates.
For a test particle, at the ISCO, $r_\mathrm{h}= 5\,(G\,m/c^2)$, so we have 
$(dr_\mathrm{h}/dt)_\mathrm{ISCO}=-(8/25)^2\,\nu\,c$. Since radial and time 
coordinates in Schwarzschild and harmonic coordinate systems are related as  
\be
\label{eq:rhrs}
r_{\rm s}=r_{\rm h}+{G\,m/c^2}
\,,\,\,\,\,\,\,\,\,\,
t_{\rm s}=t_{\rm h}=t\,.
\ee
we have for the radial velocity of the particle in Schwarzschild coordinates as
\begin{align}
\label{drsdt}
{dr_\mathrm{s}\over dt} 
&={dr_\mathrm{h}\over dt}\\\nonumber
&=-\left({8/25}\right)^2\,\nu\,c\,.
\end{align}
It is easy to show using Eq.~\eqref{eq:dtdtau} and \eqref{eq:drsdtau} that
the required energy for such an orbit will be given by
\begin{equation}
\bar{E}^2 = \frac{8}{9} \left [ 1 - \frac{9}{4}\,{1\over c^2}\left(\frac
{dr_\mathrm{s}}{dt} \right )^2_\mathrm{ISCO} \right ]^{-1}.
\label{Eisco}
\end{equation}
where ${dr_\mathrm{s}/dt}$ is given by Eq.~\eqref{drsdt}.\\

Now with $\bar{E}$ given by the above equation and the choice $\bar{L}=\sqrt{12}$, we can compute 
the desired integral numerically.\footnote{For this purpose we shall use the NIntegrate option
inbuilt in Mathematica.} As last input, for the limiting values of the integration variable $y$,
we choose $y_0=1/6$ and $y_{\rm Horizon}=(2(1+\nu))^{-1}$. Note that the choice $y_0$=1/6 
and the ones that have been made for $\bar{E}$ and $\bar{L}$ above will not be consistent with
Eq.~\eqref{eq:xy}: thus when computing the recoil velocity at the ISCO the value of the 
parameters $x$ at ISCO must be consistent with the choice for $y_0, \bar{E}$ and $\bar{L}$ made 
above.\\   

In the second method, one matches the circular orbit at the ISCO and the one associated with 
the plunge by evolving it across the ISCO. It can be performed using the energy and angular momentum
balance equations for circular orbits in the adiabatic limit at the
ISCO. For this, we shall have
\bse
\label{eq:dEJdt}
\begin{align}
\frac{d\bar{E}}{d t}&= -\frac{32}{5}\,\frac{c^3\,\nu}{G\,m}
\,x_\mathrm{ISCO}^5,\\ 
\frac{d\bar{L}}{d t} &=
{\left(G\,m\,\omega_\mathrm{ISCO}\over c^3\right)}^{-1} \,\frac{d\bar{E}}{d t} = -\frac{32}{5}\,{c^3\,\nu \over G\,m}\,x_\mathrm{ISCO}^{7/2}.
\end{align}
\ese
Following \cite{BQW05}, we can write for the quantity on the left side of Eq.~\eqref{eq:dEJdt}  
\bse
\label{dEJdt-alpha}
\begin{align}
d\bar{E}/dt=(\bar{E}-\bar{E}_\mathrm{ISCO})/(\alpha\,P),\\
d\bar{L}/dt=(\bar{L}-\bar{L}_\mathrm{ISCO})/(\alpha\,P).
\end{align} 
\ese
Here, $\alpha$ denotes a fraction of the orbital
period $P$ of the circular motion at the ISCO. Now using $\omega_\mathrm{ISCO}=({c^3/G\,m})
\,x_\mathrm{ISCO}^{3/2}$, we have for the plunge orbit
\begin{mathletters}
\label{eq:EJtilde}
\begin{eqnarray}
\bar{E} &=& \bar{E}_\mathrm{ISCO} -
\frac{64\pi}{5}\,\alpha\,\nu\,x_\mathrm{ISCO}^{7/2}\,,\\ 
\bar{L} &=&
\bar{L}_\mathrm{ISCO} -
\frac{64\pi}{5}\,\alpha \,\nu\,x_\mathrm{ISCO}^{2}\,.
\end{eqnarray}
\end{mathletters}
Finally, in the second model, in order to integrate the integral in the problem
we need to specify the limiting values for the variable $y$. For the initial value 
of the parameter (at ISCO), $y=y_0$, one can solve the following equations which 
is obtained using Eq.~\eqref{eq:xy}:  
\begin{equation}
\label{eq:y0}
x_\mathrm{ISCO} = 6^{-1} = \left[\frac{\bar{L}}{\bar{E}} \,y_0^2
(1-2y_0)\right]^{2/3}\,.
\end{equation}
For the value of the parameter, $y$, at the horizon again we take 
$y_\mathrm{Horizon}=(2(1+\nu))^{-1}$. 
For the fraction $\alpha$ of the period, we choose values between 1 and
0.01, and check the dependence of the result on this choice (see below).
\section{Results and discussions}
\label{sec:results}
\begin{figure}[t]	
\includegraphics[width=0.85\textwidth,angle=0]{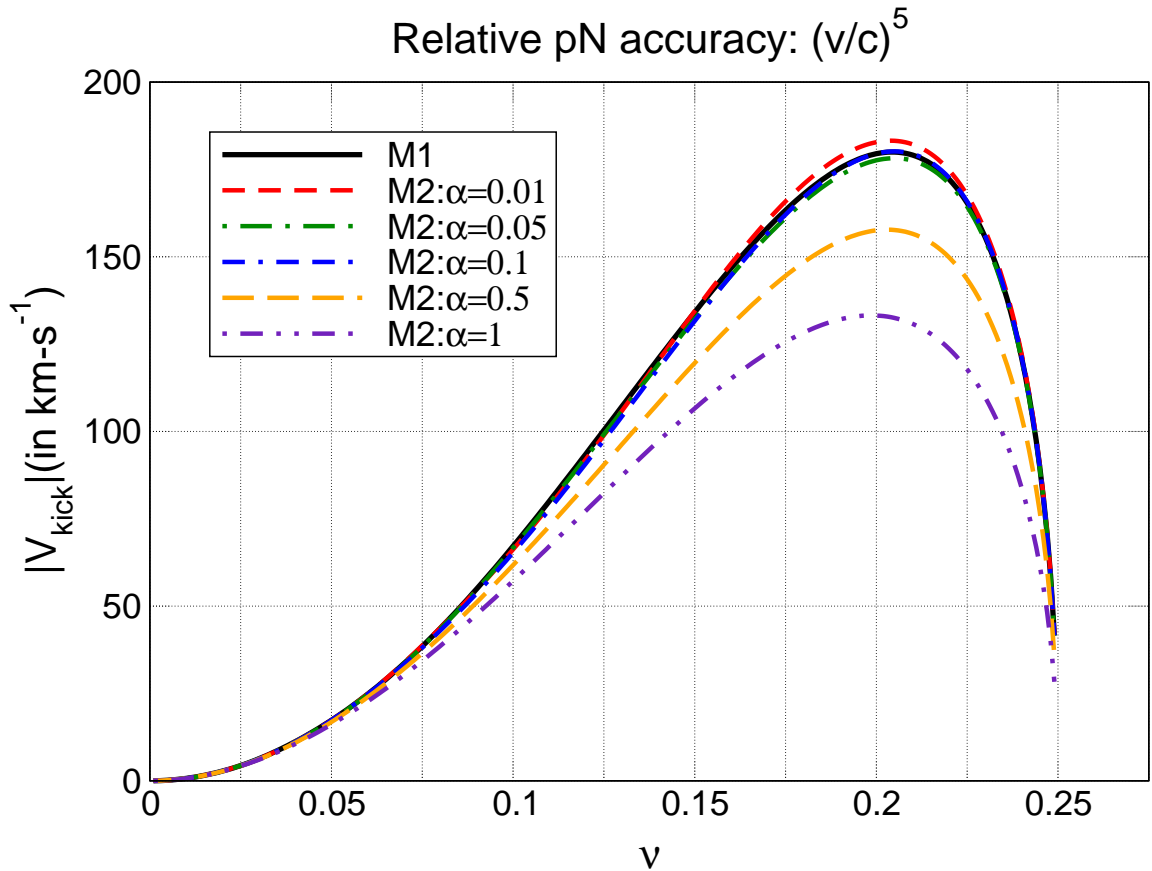}
\caption{Recoil velocity as a function of the mass parameter $\nu$ 
(symmetric mass ratio) has been shown. Plot shows a comparison between the results 
obtained using the two methods (we call them M1 and M2), discussed in 
Sec.\ref{sec:numerical-estimates}, that have been used to match the circular orbit at 
the ISCO to a suitable plunge orbit. It is evident from the figure that 
the two methods are consistent with each other for smaller values of the 
parameter $\alpha$.}
\label{fig:recvel}
\end{figure}
Equation~\eqref{eq:LMF} gives the 2.5PN formula for computing the loss rate of
linear momentum in the far-zone of a nonspinning 
inspiralling compact binary in a quasicircular orbit. In 
Sec.\ref{sec:numerical-estimates}, we show how one can numerically 
estimate the recoil velocity accumulated during the plunge phase, 
after making some simplifying assumptions. The recoil velocity at the end of 
the inspiral phase, i.e. at the fiducial ISCO, is given by Eq.~\eqref{eq:recvel-ISCO} whereas 
the recoil velocity accumulated during the plunge phase is given by 
Eq.~\eqref{eq:dVplng-y}. The integral of Eq.~\eqref{eq:dVplng-y} needs to be evaluated 
numerically keeping in mind that appropriate choices for energy and angular momentum at 
the onset of the plunge phase has been made in order to match the inspiral and plunge 
orbits at the fiducial ISCO. Figure~\ref{fig:recvel} shows our numerical estimates for the recoil 
velocity, based on the two methods (we call them M1 and M2) for matching the circular 
orbit at the fiducial ISCO to a suitable plunge orbit, discussed in the previous section. 
Figure~\ref{fig:recvel} also shows a comparison between the 
recoil velocity estimates using the two methods, M1 and M2.  
It is evident from the figure that results from both the methods 
are consistent with each other for smaller value of the parameter $\alpha$, 
defined above. In the case of M2, we have shown 
curves corresponding to $\alpha=\{0.01, 0.05, 0.1, 0.5, 1.0\}$, and one can 
see that the curves with $\alpha=0.01\,, 0.05,{\rm and}\,0.1$ are very close to 
the curve corresponding to the M1. We also observe that the recoil 
velocity, for a binary system with $\nu=0.2$, 
as shown in Fig.~\ref{fig:recvel}, is $\sim$179.5\,$\rm{km\,s^{-1}}$. 
This is lower than the 2PN accurate BQW estimate of about 
243\,$\rm{km\,s^{-1}}$ for the binary with the same mass ratio ($\nu=0.2$).
This behavior is due to the presence of large negative coefficients at the 
2.5PN order (see Eq.~\eqref{eq:recvel}) which bring down the estimates significantly. 
Such a behavior is not new to PN calculations, {\it e.g.} a similar behavior was 
observed in \cite{Wi92} at 1PN order 
(see Fig.1 of \cite{BQW05}), where the use of 1PN accurate results give a lower 
estimate for the recoil velocity as compared to the one obtained using the
Newtonian formulas since the 1PN term again contributes negatively to the recoil 
velocity. 

Note that the estimates presented in Fig.~\ref{fig:recvel} use only the 
leading order radiation reaction effects for setting initial energy (in M1) 
and energy and angular momentum (in M2). We repeat the exercise using 
2.5PN expressions for relevant quantities beyond the leading order effect 
and find that changes in estimates are negligible
(relative $\%$ changes are less than 0.5$\%$).

As discussed earlier, normally the PN approximations are expected to become less and less 
reliable beyond the ISCO. This leads to a crude estimate of the accumulated recoil velocity during 
the plunge phase. Hence, it becomes important to compare our results to some 
other numerical/analytical estimates, in order to be sure that these estimates 
are indeed reliable. In the case of the present work the closest comparison for 
the recoil velocity estimates can be made by comparing our results with those 
of BQW \cite{BQW05}. For a binary with $\nu=0.22$ and $\nu=0.23$, BQW suggest 
that the recoil velocity should lie in a range, (171-251) $\rm{km\,s^{-1}}$ 
and (146-220) $\rm{km\,s^{-1}}$, respectively. The uncertainty in their 
results has been estimated by flexing the 2PN expressions by addition of 2.5PN, 3PN and 3.5PN 
terms and then computing the maximum variation in 
their results (see \cite{BQW05} for details). Our estimates of the recoil 
velocity for a binary with $\nu=0.22$ and $\nu=0.23$ are $172\,\rm{km\,s^{-1}}$ 
and $155\,\rm{km\,s^{-1}}$, respectively, and thus our estimates lie in the window for the recoil 
velocity provided by BQW. However, we should note here that our estimates can 
also change if we add contributions coming from the 3PN and the 3.5PN terms (although 
changes may be relatively smaller). Currently, such an extension is not possible 
as we do not have sufficiently accurate inputs in order to perform such computations 
and thus it will be the subject matter of a work in the future. Our estimates 
are also consistent with an earlier numerical work \cite{Campanelli:2004zw} which 
suggests a range of values for recoil velocity between $(100-380)\,\rm{km\,s^{-1}}$ and 
$(90-290)\,\rm{km\,s^{-1}}$ for $\nu=0.22$ and $\nu=0.23$, 
respectively. As discussed in Sec.~\ref{sec:intro}, Ref. \cite{Sopuerta:2006et} 
suggests that maximum recoil velocity estimate for a binary with $\nu=0.2$ in quasicircular 
orbit lie in a range between (79-216) $\rm{km\,s^{-1}}$. As mentioned above, our estimate for 
such system is $\sim$179.5\,$\rm{km\,s^{-1}}$ and thus is consistent with their estimates. 

We witnessed above that inclusion of 2.5PN contributions significantly changed 
earlier PN estimates for the recoil velocity indicating that contributions at higher 
orders need to be explicitly assessed due to the asymptotic nature of the PN 
expansion. As mentioned above, contributions at other high PN orders 
such as at 3PN and 3.5PN should be included in some future work in order 
to have better estimates for the recoil velocity, although the changes may be relatively 
smaller as compared to those brought in by 2.5PN contributions. A numerical study 
\cite{Baker:2006vn} suggests that the recoil velocity estimates at the fiducial ISCO should be  
of the order of $\sim$ 14\,$\rm{km\,s^{-1}}$ for a binary with $\nu=0.24$ and this estimate matches 
well with BQW estimates for the same system. This is a relatively higher estimate as compared to our 
estimate of 2.8\,$\rm{km\,s^{-1}}$ at the fiducial ISCO for a system with the same mass ratio. In such a case, 
we should 
expect that inclusion of higher order contributions at the 3PN order will contribute 
to the recoil velocity positively (in contrast to the negative contributions from 2.5PN terms) 
and thus could bring up the 
estimates to match with estimates of \cite{Baker:2006vn} and BQW. 
In addition to this, 
as a follow-up of this work, 
one can try to include contributions due to the final ringdown phase using 2.5PN 
accurate initial conditions\footnote{One can follow \cite{LeTiec:2009yg}, where a method 
of computing the contribution due to ringdown phase was proposed and used 2PN initial 
conditions which were obtained in BQW.} and then combine this with the recoil velocity estimates 
for the inspiral and plunge phase presented here. 
This will allow one to make more direct 
comparisons with the results obtained using numerical relativity and
the effective one-body approach which include contributions from all three 
phases of the binary evolution. 
\acknowledgments
We thank Luc Blanchet for useful discussions. KGA acknowledges the hospitality 
of Raman Research Institute  at various stages during the project. CKM acknowledges the hospitality of the Chennai 
Mathematical Institute during Fall 2010. KGA acknowledges discussions with M S S Qusailah during the initial phase 
of the project. 
\bibliography{ref-list}
\end{document}